\documentclass[aps,pra,letterpaper,10pt,twocolumn,superscriptaddress]{revtex4-1}
\usepackage[colorlinks=true,allcolors=blue]{hyperref}
\usepackage{amssymb}
\usepackage{amsmath}
\usepackage{graphicx}
\usepackage[caption=false]{subfig}
\usepackage{amsfonts}
\usepackage{xcolor}
\usepackage{epsfig}
\usepackage{color}
\usepackage{bm}
\usepackage{tabularx}
\usepackage{multirow}
\usepackage{mathtools}
\usepackage{xfrac}
\usepackage{blkarray}
\usepackage{bbold}
\usepackage[mathscr]{euscript}
\usepackage{autobreak}
\usepackage{comment}

\begin{document}
\title{Exceptionally strong coupling of defect emission in hexagonal boron nitride to stacking sequences}

\author{Song Li}
\thanks{These authors contributed equally to this work.}
\affiliation{Wigner Research Centre for Physics, P.O.\ Box 49, H-1525 Budapest, Hungary}

\author{Anton Pershin}
\thanks{These authors contributed equally to this work.}
\affiliation{Wigner Research Centre for Physics, P.O.\ Box 49, H-1525 Budapest, Hungary}
\affiliation{Department of Atomic Physics, Institute of Physics, Budapest University of Technology and Economics,  M\H{u}egyetem rakpart 3., H-1111 Budapest, Hungary}

\author{Pei Li}
\affiliation{Beijing Computational Science Research Center, Beijing 100193, China}
\affiliation{Wigner Research Centre for Physics, P.O.\ Box 49, H-1525 Budapest, Hungary}

\author{Adam Gali}
\email{gali.adam@wigner.hu}
\affiliation{Wigner Research Centre for Physics, P.O.\ Box 49, H-1525 Budapest, Hungary}
\affiliation{Department of Atomic Physics, Institute of Physics, Budapest University of Technology and Economics,  M\H{u}egyetem rakpart 3., H-1111 Budapest, Hungary}

\date{\today}
\begin{abstract}
Van der Waals structures present a unique opportunity for tailoring material interfaces and integrating photonic functionalities. By precisely manipulating the twist angle and stacking sequences, it is possible to elegantly tune and functionalize the electronic and optical properties of layered van der Waals structures. Among these materials, two-dimensional hexagonal boron nitride (hBN) stands out for its remarkable optical properties and wide band gap, making it a promising host for solid state single photon emitters at room temperature. Previous investigations have demonstrated the observation of bright single photon emission in hBN across a wide range of wavelengths. In this study, we unveil an application of van der Waals technology in modulating their spectral shapes and brightness by carefully controlling the stacking sequences and polytypes. Our theoretical analysis reveals remarkably large variations in the Huang-Rhys factors—an indicator of the interaction between a defect and its surrounding lattice—reaching up to a factor of 3.3 for the same defect in different stackings. We provide insights into the underlying mechanism behind these variations, shedding light on the design principles necessary to achieve rational and precise control of defect emission. This work paves the way for enhancing defect identification and facilitating the engineering of highly efficient single photon sources and qubits using van der Waals materials.  

\end{abstract}

\maketitle

%
%
\section{Introduction}
In recent developments in the field of two-dimensional materials, nanodevices utilizing graphene and hexagonal boron nitride have undergone a unique evolution. Initially, graphene served as the active material while hBN functioned solely as a substrate or capping layer due to its excellent chemical stability ~\cite{dean2010boron, yang2013epitaxial, petrone2015flexible}. However, the roles have now reversed, with graphene primarily acting as an electrode while hBN has emerged as the active material for applications in optoelectronics, quantum optics, and quantum information science. The intrinsic properties of hBN have become a topic of great interest, including its potential for field-enhanced molecular sensing through strong coupling to molecular vibrations~\cite{pons2019launching, dai2014tunable}, as well as its ability to host room-temperature magnetic textures when interfaced with metallic ferromagnets~\cite{el2023evidence}.
Furthermore, the large band gap of hBN can accommodate the localised levels from the deep point defects, providing a platform for the solid-state spin qubits with optically addressable spin states ~\cite{gottscholl2020initialization, chejanovsky2021single, mendelson2021identifying, stern2022room, liu2022spin,vaidya2023quantum}. These spin qubits can be applied for quantum sensing~\cite{gottscholl2021spin, yang2022spin, gao2023quantum}. To date, numerous defect-related single photon emitters (SPE) in hBN have been reported, covering a wide range of wavelengths from infrared to ultraviolet~\cite{tran2016quantum, tran2016robust, tran2016quantum, bourrellier2016bright, chejanovsky2021single}. These SPEs hold great potential for quantum information processing, but their precise origin should still be identified to enable the realization of these applications.

Dozens of defect structures in hBN, such as native defects, and carbon and oxygen impurities, to name a few, have been proposed as sources of SPEs~\cite{mendelson2021identifying, chejanovsky2021single, li2022identification, hamdi2020stone, weston2018native, li2022carbon, jara2021first, sajid2020vncb, Gao2021}. However, achieving the coherent control of the spin state has mostly been demonstrated with boron vacancy~\cite{liu2022spin,gottscholl2021room,Haykal2022,liu2022coherent}, the only spin defect that was unambiguously identified in hBN~\cite{gottscholl2020initialization,ivady2020ab,sajid2020edge}. The lack of assigned defect structures is rooted in substantial variations of photoluminescence signals, which serve as the primary means of identification, across different samples of the material. While the local strain effects are commonly attributed to this behaviour, the impact of other intrinsic two-dimensional phenomena, such as twisting, sliding, and variations in layer stacking~\cite{vizner2021interfacial, yasuda2021stacking, yao2021enhanced, wilson2021excitons, caldwell2019photonics, wang2022interfacial, woods2021charge, ochoa2020flat, zhao2020formation}, on electronic properties of defects remains poorly understood. Sliding, for instance, occurs across a monolayer step, leading to a gradual change between two stacking patterns~\cite{woods2021charge}.
The twist angle can manipulate the strength of phonon-phonon coupling~\cite{ouyang2020controllable}, thereby influencing the design of SPEs with desirable sharp emission lines. There is compelling evidence that the properties of hBN are closely tied to its stacking sequence. As such, interfacial ferroelectricity in hBN has been reported, with electric polarization depending on the stacking order~\cite{yasuda2021stacking, vizner2021interfacial}. Flat bands have been the subject of several theoretical reports, revealing a splitting of the band edge states induced by different stacking patterns~\cite{ochoa2020flat, zhao2020formation}.
In the realm of defect emission, recent studies have shown that the zero phonon line (ZPL) of ultraviolet emission from defects is stacking-dependent~\cite{rousseau2022bernal, rousseau2022stacking}. Moreover, the brightness of this emission can be greatly enhanced by twisting the hBN layers and further tuned by an external electric field~\cite{su2022tuning}. These intriguing findings serve as the motivation for our study, wherein we explore the effects of stacking and sliding on deep level emission in hBN.

This paper presents comprehensive theoretical calculations for the optical properties of common ultraviolet SPEs in hBN with a specific focus on different stacking sequences. Our results demonstrate a shift of ZPL of 2C defect from $\text{AA}^{\prime}$ to AB stacking, which is consistent with experimental data indicating ZPL energies of 4.09 eV for $\text{AA}^{\prime}$ stacking and 4.14 or 4.16 eV for Bernal (AB) stacking~\cite{rousseau2022stacking}. Most strikingly, the choice of stacking sequences also affects the shape of photoluminescence  spectrum  by altering the PL sideband. This phenomenon can be attributed to variations in the interlayer electrostatic Coulomb interactions. Our calculations indicate that the optical lifetime of common defects in the regular stacking patterns is primarily determined by the in-plane components of the transition dipole moment. Coupled with the inversion symmetry observed along the out-of-plane direction, this property acts as a protective factor, shielding the defects from the impact of out-of-plane electric fields. However, when situated in an inhomogeneous environment from misaligned layers of hBN, the dipolar defects demonstrate a tendency to align themselves with the direction of the field. This behavior becomes most evident in twisted bilayers, where we observe a substantial alteration in the photoluminescence signals, in agreement with experimental observations.~\cite{su2022tuning}.  Our findings provide useful insights into stacking-dependent emission from defects in hBN and offer a unique strategy to enhance the brightness and quantum efficiency of SPEs. 

\begin{figure}[tb]
\includegraphics[width=0.9\columnwidth]{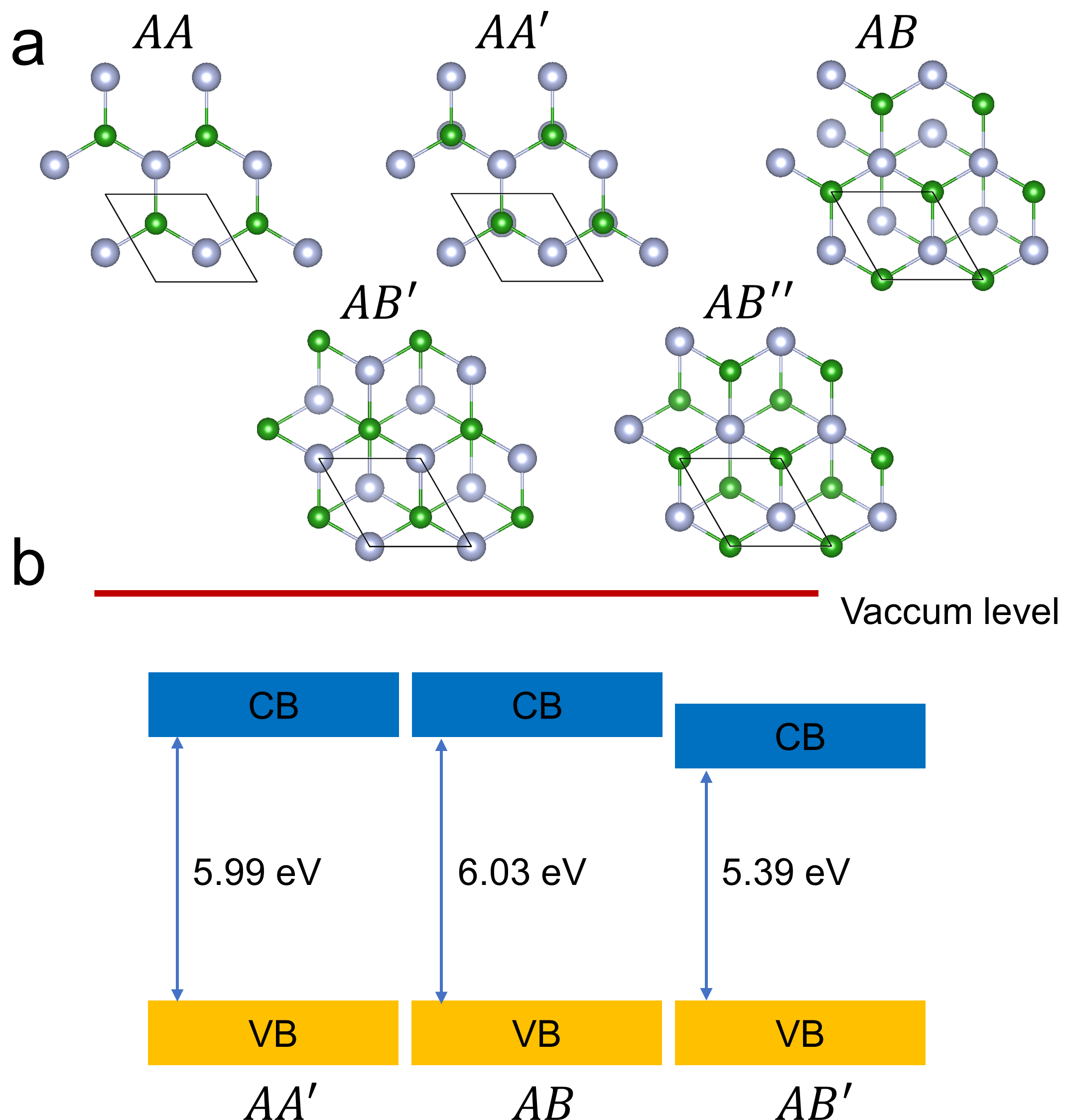}
\caption{\label{Figure1}%
(a) Schematics view of the five high-symmetry stackings in bilayer hBN. In the AA stacking, boron (nitrogen) atoms in the bottom layers are fully aligned with boron (nitrogen) atoms in the top layer. The $\text{AA}^{\prime}$ stacking is essentially similar to AA, but the boron atoms are aligned with nitrogen atoms. The AB is formed by rotating the layers of the $\text{AA}^{\prime}$ stacking by 60 degrees. The difference between the $\text{AB}^{\prime}$ and $\text{AB}^{\prime\prime}$ stacking is that either nitrogen or boron atoms appear in the center of the honeycomb from another layer. (b) Band alignment diagrams for the stacking sequences from (a) showing the position of the band edges relative to the vacuum level and computed for the representative slab models. The respective band gap energies in bulk are 6.05 eV, 6.06 eV and 5.30 eV}
\end{figure}

\begin{figure}[tb]
\includegraphics[width=\columnwidth]{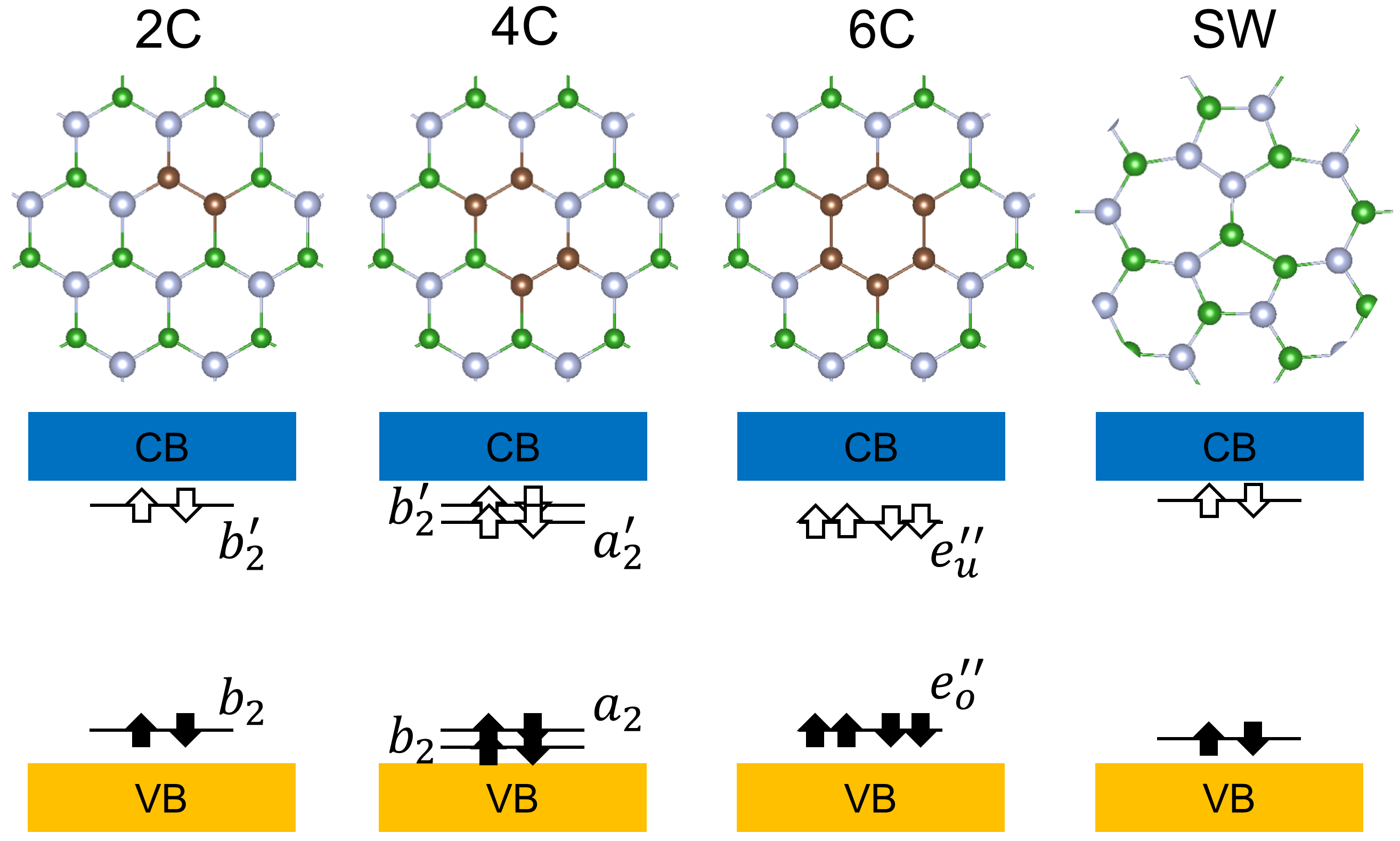}
\caption{\label{Figure2}%
Schematics view of the three carbon clusters  and a SW defect in hBN which we consider as possible SPE sources in the ultraviolet range. The bottom panels show the defect levels inside the band gap of hBN, obtained from the ground state HSE calculations.}
\end{figure}

\section{Results}
\label{sec:results}

According to the stacking order of the successive sheets or layers, a bilayer of hBN can emerge in five different high-symmetry stackings~\cite{gilbert2019alternative,constantinescu2013stacking}, as shown in Figure~\ref{Figure1}a. First, we compared the relative stability of these sequences by computing the total energies and checking for the appearance of imaginary modes in the phonon calculations. This analysis revealed that the AA and $\text{AB}^{\prime\prime}$ stackings are unstable, as indicated by the presence of imaginary phonon modes at 10.4 meV and 8.3 meV, respectively. These results are consistent with a previous study that reported higher total energies for these two stacking sequences ~\cite{gilbert2019alternative}.
In fact, the $\text{AA}^{\prime}$ stacking pattern is commonly observed in most synthesized hBN samples, and its properties have been extensively studied in the past decades~\cite{bourrellier2014nanometric, kim2015synthesis, alem2009atomically}. Interestingly, we found that the AB stacking has a lower total energy than the conventional $\text{AA}^{\prime}$ stacking, and its presence has been confirmed by HR-TEM imaging and by combining second harmonic generation (SHG) with photoluminescence spectroscopy ~\cite{warner2010atomic, rousseau2022bernal}. Therefore, we focus our attention on the three stable polytypes, namely, $\text{AA}^{\prime}$, AB and $\text{AB}^{\prime}$.  Our calculations were able to reproduce the experimental band gap of approximately 6.1 eV for $\text{AA}^{\prime}$ and AB. Note that this value neglects a contribution of the zero-point renormalization, which arises from electron-phonon interactions~\cite{cassabois2016hexagonal}. We also observe a substantial decrease in the band gap energy to 5.3 eV for $\text{AB}^{\prime}$, as shown in Figure~\ref{Figure1}b. Aligned with the vacuum level, it becomes evident that the decrease in the band gap of $\text{AB}^{\prime}$ is primarily attributed to the shift in the conduction band minima (CBM), while the valence band maximum (VBM) remains unchanged.    

\begin{figure}
\includegraphics[width=\columnwidth]{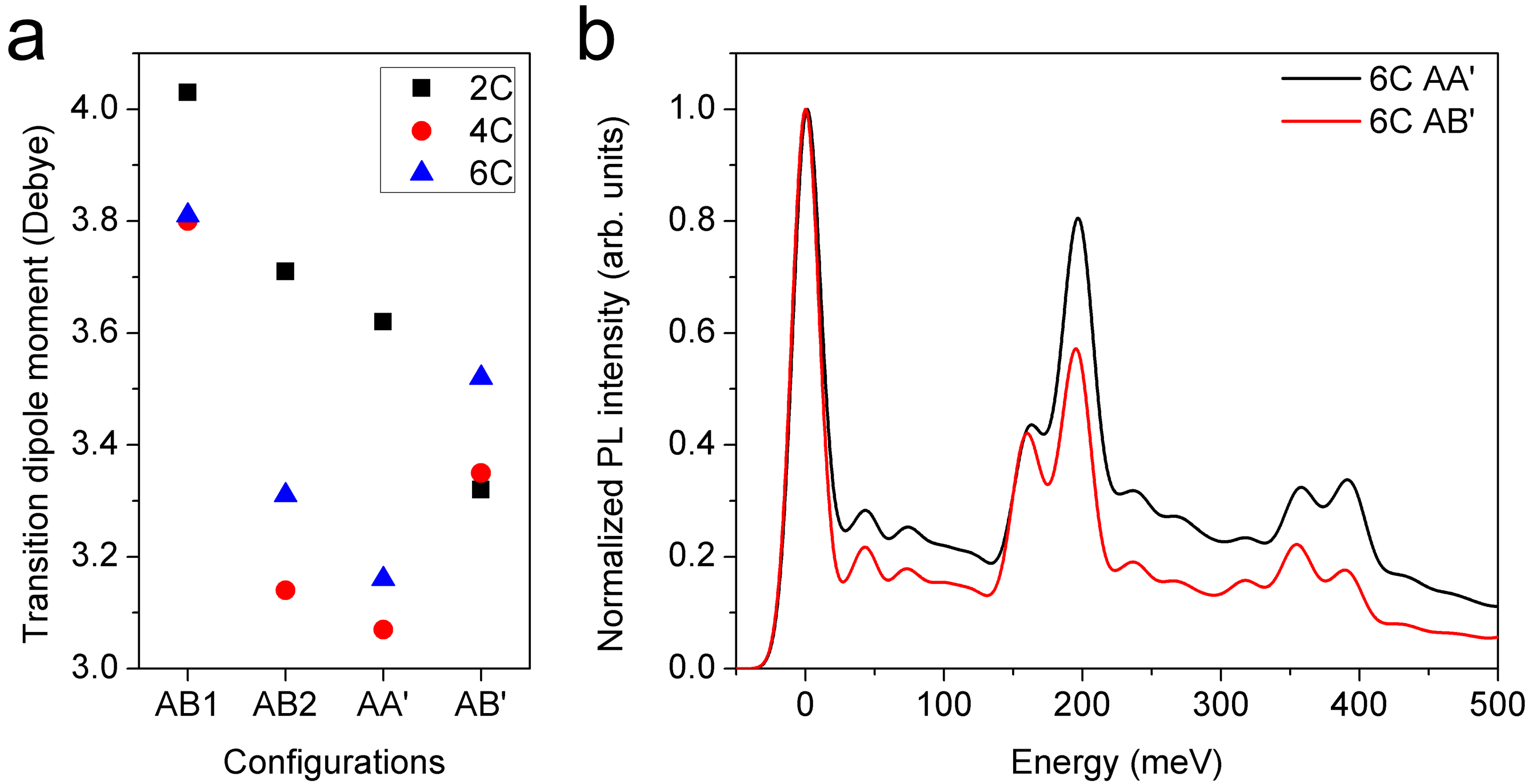}
\caption{\label{Figure3}%
(a) Variations of transition dipole moments calculated for the three carbon defects in the different stackings of hBN. (b) The simulated PL spectra of 6C defects in the $\text{AA}^{\prime}$ and $\text{AB}^{\prime}$ stacking sequences.}
\end{figure}

In the literature, several compelling models exist to accurately describe the defects responsible for single-photon emission in the ultraviolet region of hBN. These models include a 2C (C$_{\rm N}$C$_{\rm B}$) defect ~\cite{mackoit2019carbon}, as well as our recent findings on carbon complexes, and a Stone-Wales (SW) defect ~\cite{hamdi2020stone, li2022ultraviolet}. The structures and energy diagrams of these defects are shown in Figure~\ref{Figure2}. All these defects maintain a stable neutral charge state over an energy range that exceeds the ionization threshold. The 2C defect exhibits one occupied and one empty state within the band gap, both possessing a $b_2$ symmetry. The occupied state primarily originates from a $p_z$ orbital of C$_{\rm N}$, while the empty state arises from a $p_z$ orbital of C$_{\rm B}$. On the other hand, the 4C defect features two additional levels within the energy gap between the $b_2$ states, resulting in the lowest optical transition from $a_2$ to $a^{\prime}_2$. In contrast to the previous defects, the 6C carbon ring exhibits a $D_{3h}$ symmetry, giving rise to degenerate e states within the band gap. These states are labeled as occupied $e^{\prime\prime}_o$ and unoccupied $e^{\prime\prime}_u$. Similar to the 2C defect, the SW defect also possesses one occupied and one empty defect level. Despite occasional variations in the absolute energies, the basic electronic structure of these defects remains well-preserved across different stackings. In Supplementary Fig. 2, we provide detailed defect configurations that we modeled in the three polytypes. It is worth noting that we considered two nonequivalent lattice sites for AB stacking, denoted as AB1 and AB2. In AB1 stacking, carbon atoms are aligned with nitrogen atoms, whereas in AB2 stacking, carbon atoms are aligned with boron atoms.

We present the calculated parameters, including ZPLs, Huang-Rhys (HR) factors, and radiative lifetimes, in Table~\ref{tab:data1}. Our results show that compared to the previous work on 2C ~\cite{mackoit2019carbon}, our calculations yield lower ZPLs and longer radiative lifetimes. This difference can be attributed to the varying fraction of the Fock exchange. Importantly, we account for the two-determinant nature of the excited singlet state in our ZPL calculations through a correction term, as discussed in Supplementary Note 2.
Our calculations reveal that interlayer interaction greatly impacts the photoluminescence spectrum. First, for all three carbon defects, the ZPLs increase when transitioning from $\text{AA}^{\prime}$ to AB stacking, while the variations are found to be system-specific. In particular, 2C shows a small change in ZPL, from 4.09 eV in $\text{AA}^{\prime}$ stacking to 4.24 or 4.21 eV with AB stacking in accordance with experimental observations~\cite{rousseau2022stacking}. Notably, despite the substantially smaller band gap in $\text{AB}^{\prime}$ stacking, the ZPLs only experience a slight shift. Furthermore, the stacking arrangement affects the relaxation energy in the excited state. For the 2C defect, the relaxation energy is 0.21 eV and 0.24 eV for the AB1 and AB2 configurations, respectively. This suggests a relatively stronger electron-phonon coupling with the AB1 pattern and highlights the influence of stacking on the optical transitions of defects. This effect is further demonstrated by the calculated HR factors, which increase from 1.80 to 2.02 when transitioning from AB1-2C to AB2-2C. By contrast, the changes in lifetime range from 1.5 to 3.3 ns, which, despite notable variations in the transition dipole moment shown in Figure~\ref{Figure3}a, are relatively small. This is because the increase in the dipole moment is offset by the decrease in ZPL energies. Thus, the primary effect of the stacking arrangements is the modification of the   photoluminescence shape, with the narrowest signals observed in $\text{AB}^{\prime}$ stacking. To further illustrate the concept of sideband engineering by altering the stacking sequence, Fig. ~\ref{Figure3}b depicts the PL sidebands of $\text{AA}^{\prime}$ and $\text{AB}^{\prime}$ for the 6C defect.

\begin{figure}
    \includegraphics[width=\columnwidth]{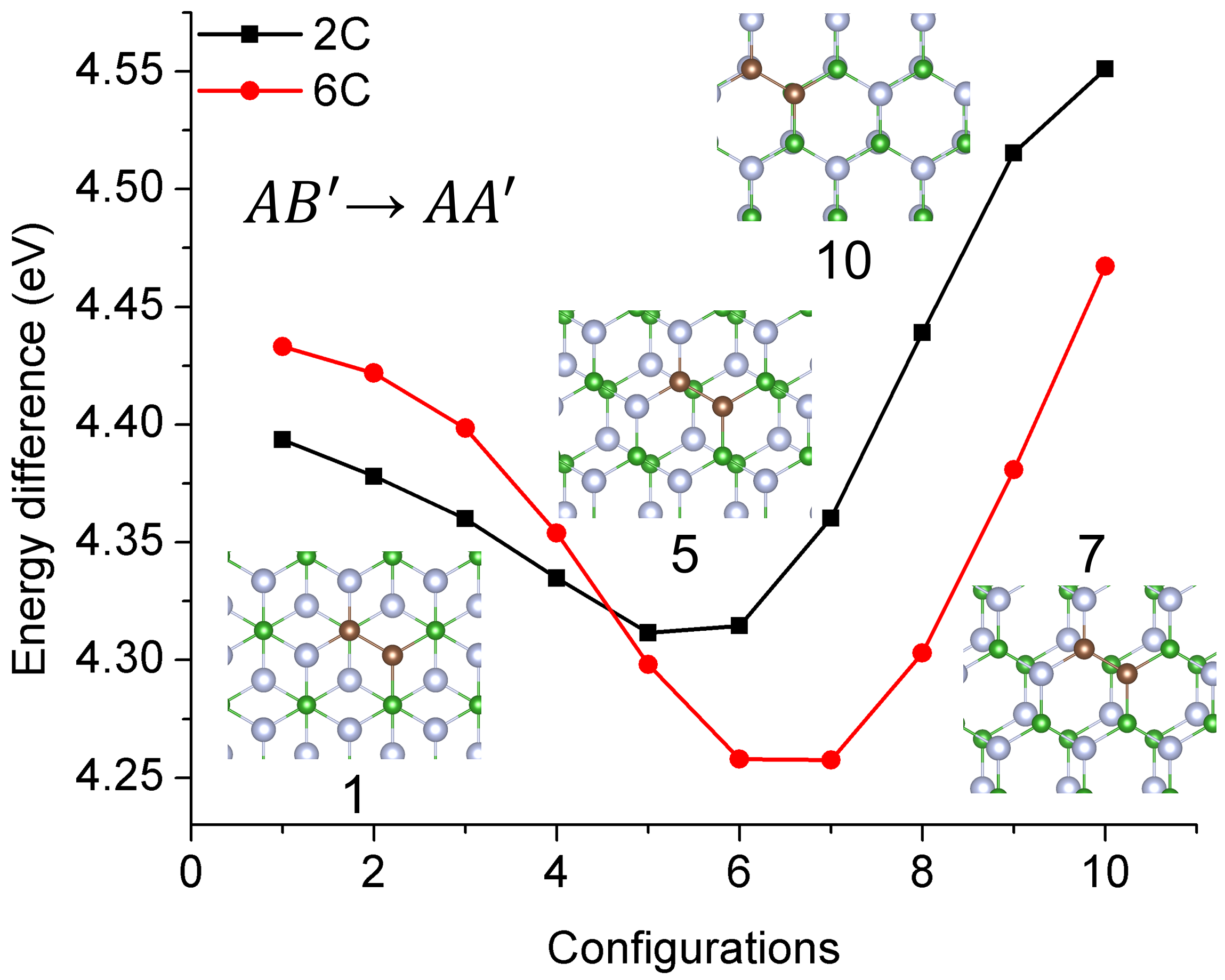}
     \caption{\label{Figure4}The calculated energy difference between the highest occupied and the lowest unoccupied defect levels of 2C and 6C as a function of sliding. Here, the configuration 1 corresponds to the $\text{AB}^{\prime}$ stacking, while configuration 10 denotes the $\text{AA}^{\prime}$ stacking. The insets show the respective configurations of the 2C defect along the sliding path.}
\end{figure}


\renewcommand{\arraystretch}{1.5}
\begin{table}[htb]
\caption{\label{tab:data1} Calculated ZPL energies after including the correction term (eV), HR factors, transition dipole moments (Debye) and radiative lifetimes (ns) for the carbon impurities and SW defect in the different stackings of hBN.}
\begin{ruledtabular}
\begin{tabular}{l|cccc}
Defect   & ZPL  & HR factor & TDM  & Lifetime \\
\hline
$\text{AA}^{\prime}$-2C & 4.09 & 1.94 & 3.62 & 2.59\\
$\text{AB1}$-2C & 4.24 & 1.80 & 4.03 & 1.88\\
 $\text{AB2}$-2C & 4.21 & 2.02 & 3.71 & 2.28\\
$\text{AB}^{\prime}$-2C & 4.00 & 1.68 & 3.32 & 3.32\\
\hline
$\text{AA}^{\prime}$-4C & 4.32 & 1.47 & 3.07 & 3.07\\
$\text{AB1}$-4C & 4.56 & 1.20 & 3.80 & 1.47\\
$\text{AB2}$-4C & 4.45 & 1.58 & 3.14 & 2.39\\
$\text{AB}^{\prime}$-4C & 4.28 & 1.10 & 3.35 & 2.66\\
\hline
$\text{AA}^{\prime}$-6C & 4.17 & 2.09 & 3.16 & 3.22\\
$\text{AB1}$-6C & 4.49 & 1.74 & 3.81 & 1.85\\
$\text{AB2}$-6C & 4.31 & 1.96 & 3.31 & 2.67\\
$\text{AB}^{\prime}$-6C & 4.22 & 1.56 & 3.52 & 2.34\\
\hline
$\text{AA}^{\prime}$-SW & 3.99 & 3.12 & 1.59 & 14.44\\
$\text{AB1}$-SW & 4.01 & 3.21 & 1.56 & 14.93\\
$\text{AB2}$-SW & 3.97 & 3.62 & 1.66 & 13.29\\
$\text{AB}^{\prime}$-SW & 3.97 & 2.94 & 1.85 & 10.92\\

\end{tabular}
\end{ruledtabular}
\end{table}

We proceed to investigate the effect of sliding which is feasible between the $\text{AA}^{\prime}$ and $\text{AB}^{\prime}$ configurations. Due to the rapid recovery of a high symmetry configuration (either $\text{AA}^{\prime}$ or $\text{AB}^{\prime}$), performing a full geometry optimization of the defected structures becomes cumbersome. Therefore, our focus is on the energy difference between defect levels for the sliding from $\text{AB}^{\prime}$ to $\text{AA}^{\prime}$ configurations, as illustrated in Fig.~\ref{Figure4}. We observe a decrease in the energy difference when the geometry is in an off-high-symmetry configuration, reaching a value within 0.25 eV. This observation suggests that the ZPLs may shift during the sliding process, as well. The evolution of the energy difference is primarily driven by the changes in the band gap of pristine hBN during sliding, as depicted in Supplementary Fig. 9. Generally, the band gap decreases with non-centrosymmetric stackings, reaching a minimum when the other layer lies on a bridge site. There is an explanation for these effects, which suggests that chemical (Coulomb) interactions play a crucial role in determining the relative stability of different stackings, while electron correlation softens the potential energy surface~\cite{constantinescu2013stacking}. 

\begin{figure}
    \includegraphics[width=\columnwidth]{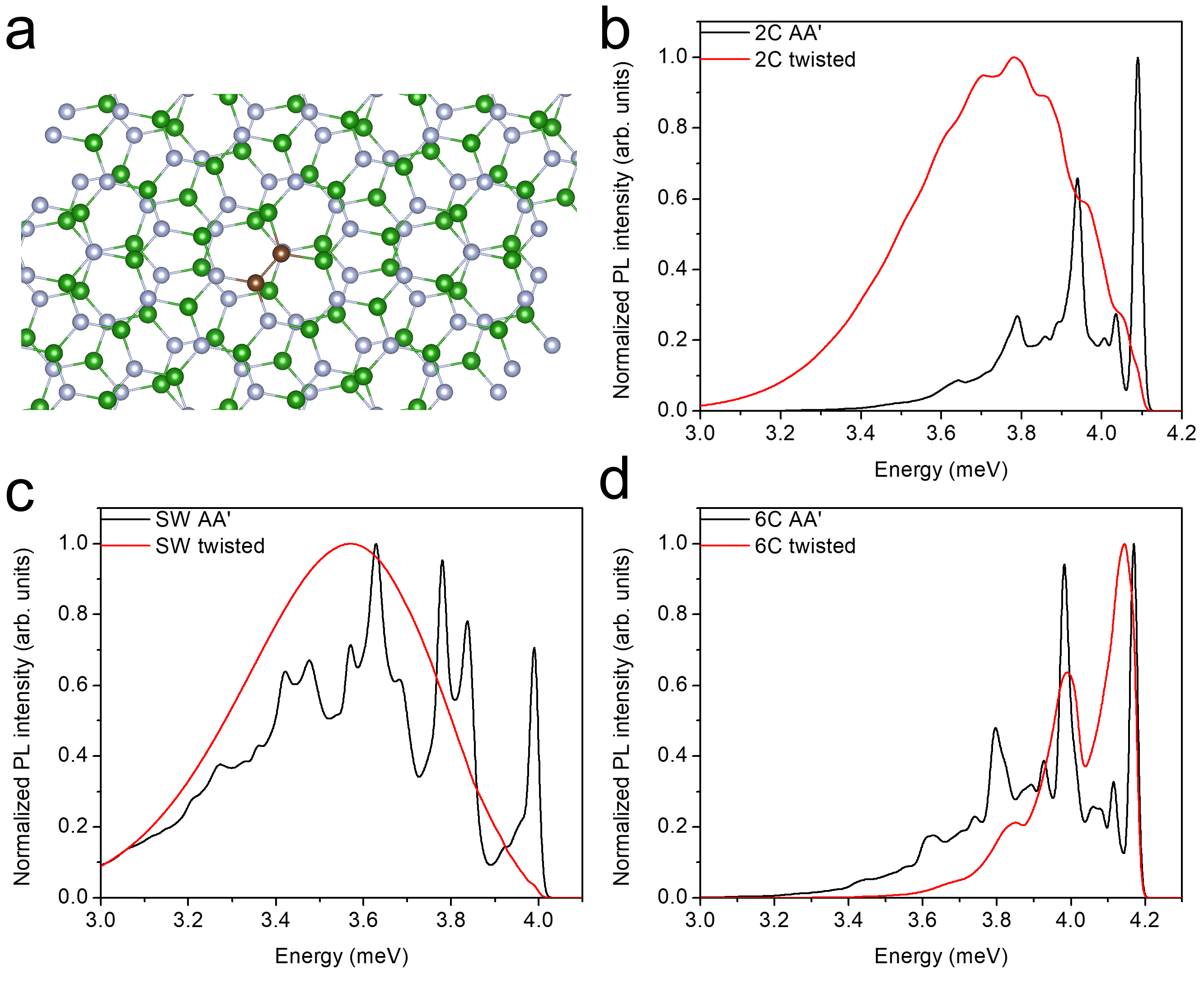}
     \caption{\label{Figure5} (a) The structure of the 2C defect in the twisted bilayer of hBN with the twist angle of 21.79$^{\circ}$ (b-d) The PL spectra simulated for the  2C, SW, and 6C defects, respectively, in the twisted bilayer of hBN. For the sake of comparison, the respective spectra are in $\text{AA}^{\prime}$ bulk are also provided.}
\end{figure}

Having described the properties of single-photon emitters in different polytypes, we now shift our focus to the defect-related photoluminescence observed in twisted hBN bilayers. Unlike symmetric translations, the locally inhomogeneous environment resulting from twisting induces an out-of-plane net field. This phenomenon, in turn, can interact with the dipole moments of both ground and excited states, consequently modifying the emission spectra (also known as the Stark effect). More precisely, we examine the influence of these fields on the photoluminescence properties of 6C, 2C, and SW defects. In Fig.~\ref{Figure5}, we compare their PL sidebands calculated in the twisted bilayers and $\text{AA}^{\prime}$ stacking. It becomes evident that the response of these defects to twisting differs remarkably. However, this effect can be understood by analyzing the computed changes in dipole moments in the ground and excited states \cite{bilot1962theorie}. As confirmed by the results in Table~\ref{tab:data2}, the degree of correlation between the changes in dipole moment and the HR factors is large, given that a complete relaxation toward the net fields is impeded by the hBN lattice. Specifically, the response of the 6C defect to twisting only exhibits marginal variation, while the magnitude of changes falls within the range of the effect observed in different stacking sequences. On the other hand, striking modifications are observed for the 2C and SW defects, largely due to their substantial variations in dipole moments upon excitation. It is worth noting that, due to the change in dielectric environment, the positions of the zero-phonon line remain essentially stable for each defect.

\renewcommand{\arraystretch}{1.5}
\begin{table}[htb]
\caption{\label{tab:data2} Calculated changes in the dipole moments upon excitation (Debye) and HR factors, computed for the given defects in the $\text{AA}^{\prime}$ stacking and twisted bilayers. For 2C and SW, the (vertical) $\Delta\mu$ was computed by the TDDFT approach. For 6C, the TDDFT results in a zero-value of $\Delta\mu$; yet we assume a finite $\Delta\mu$ due to the missing contribution of the product Jahn-Teller effect \cite{li2022ultraviolet}.}
\begin{ruledtabular}
\begin{tabular}{l|ccc}
Defect   & HR ($\text{AA}^{\prime}$)  & HR (twisted) & $\Delta\mu$ \\
\hline
6C & 2.08 & 2.69 & $\sim$0.1 \\
2C & 1.94 & 5.60 & 0.6 \\
SW & 3.12 & 6.94 & 5.5 \\

\end{tabular}
\end{ruledtabular}
\end{table}

Among these defect configurations, the response of the 2C defect is particularly intriguing, as its sideband modifications align remarkably well with experimental measurements (see Supplementary Fig. 8b). Additionally, in the relaxed excited state geometry, we observe a distortion of the defect axis out of the basal plane by approximately 9 degrees. For the sake of reference, this distortion is approximately 3 degrees in the $\text{AA}^{\prime}$ bilayer and 0 degrees in bulk. The out-of-plane distortion may have a dual effect on the observed photoluminescence intensity. Firstly, due to its alignment with the direction of the laser beam in a typical optical setup, a permanent component of the transition dipole moment ($d_z$) enhances both absorption and emission. Secondly, further modifications of the wavefunctions can result from an additional geometrical relaxation. By discriminating the contributions of these two effects, we obtain the enhancement of the PL intensity from the $d_z$ contribution by a factor of 8.9, once again aligning with experimental measurements. In turn, the total PL intensity remains rather stable suggesting that the observed effect is primary caused by the defect reorientation toward the net field. Therefore, based on a full agreement with the experimental results, we put forward the 2C defect as the primary source of the experimental response of the 4.1-eV emission observed in the twisted bilayers of hBN.

It is important to note that Su et al.~\cite{su2022tuning} provided a different theoretical explanation for the modification of the optical signal in the twisted bilayers. However, despite utilizing an advanced electronic structure method, the authors did not consider the reorganisation of the ions consisting of the defects upon excitation. On the other hand, our calculations reveal the dominant role of the Stark effect, which appears to significantly impact the identification of defects in hBN. Moreover, these calculations offer new insights for experimentally validating the proposed mechanism. As depicted in Supplementary Fig.~8a, the interlayer twist activates vibrational modes within the energy range of $\sim$10-100 meV. Specifically, the modes at close to 20 meV are responsible for the out-of-plane distortion. We believe that these new signals could be observable in resonance Raman measurements and would further confirm the environmental modulation of intra-defect emission in hBN.
 
%
%
\section{Summary and conclusion}
\label{sec:summary}

In summary, we propose modifying the photoluminescence response of single photon emitters in hexagonal boron nitride by altering the stacking sequences. Our calculations indicate remarkable changes in the HR factors, with variations of up to 50\% observed for certain defects with regular polytypes. The dipolar defects exhibit strong coupling to the polytype, indicating a prominent role of the Stark effect. Given the general interest in SPEs with sharp emission, the $\text{AB}^{\prime}$ stacking is expected to produce the narrowest PL signals. By introducing twisting, the effect can be further enhanced, leading to a complete transformation of the sideband shape and effectively increased brightness. Therefore, when comparing calculated spectra to experimental data, caution should be exercised given the remarkable impact of PL spectroscopy on defect identification. Our calculations for the 2C defect align particularly well with the available experimental data, suggesting that the 2C defect is likely the primary source of the 4.1 eV-emission. In turn, its exceptional signal variations could be exploited to monitor local rearrangements of hBN caused by stain, electric fields, and other perturbations. The current technique also improves the capability of photoluminescence  measurements, enabling more effective identification of defect structures at large. Future investigations should focus on comprehending the coupling between defect spin properties and environmental changes in the different arrangements of hBN layers.

\section{Methods} 
\label{sec:methods}

Our density functional theory (DFT) calculations were performed by the \textit{Vienna ab initio simulation package} (VASP) code~\cite{kresse1996efficiency, kresse1996efficient} using a plane wave basis. Projector augmented wave (PAW) potentials~\cite{blochl1994projector, kresse1999ultrasoft} were used with a cutoff energy of 450~eV. A $9\times9$ two-layered supercell model was constructed to avoid the interactions of defect with its periodic images and to apply the $\Gamma$-point sampling scheme. The interlayer vdW interaction was described with DFT-D3 method of Grimme  ~\cite{grimme2010consistent} for the dispersion correction. To accurately account for the band gap energy, we modified the screened hybrid density functional of Heyd, Scuseria, and Ernzerhof (HSE) ~\cite{heyd2003hybrid} with a mixing parameter $\alpha = 0.32$ for the Fock exchange. The geometry optimization and calculation of electronic properties were performed with the HSE functional in consistency with our previous studies. The convergence threshold for the forces was set to 0.01~eV/\AA.  $\Delta$SCF method~\cite{gali2009theory} was used to calculate excited states. Since the interlayer distance of different stackings does not change significantly, we fix it in our model to the value of 3.31 \AA. In addition, we constructed a bilayer configuration of 252 atoms to investigate the impact of twisting on the properties of SPE.  To avoid lattice mismatch between the layers~\cite{shallcross2008quantum}, we selected a twist angle of 21.79$^{\circ}$ which falls into a region of intensified PL signal for the 4.1 eV-emission ~\cite{su2022tuning}.

The PL spectrum was simulated using the Franck-Condon approximation by computing the overlap between the phonon modes in the ground and excited states.~\cite{gali2009theory,alkauskas2014first}. The phonon modes were calculated with the Perdew-Burke-Ernzerhof~\cite{perdew1996generalized} (PBE) functional, which is a widely used, reliable and time-saving approximation. The radiative lifetime was computed using the following equation:
\begin{equation}\label{eq:1}
\Gamma_\text{rad} =\frac{1}{\tau_\text{rad}} =  \frac{n_DE^3_\text{ZPL}\mu^2}{3\pi\epsilon_{0}c^3\hbar^4}\text{,}
\end{equation}
where $\epsilon_0$ is the vacuum permittivity, $\hbar$ is the reduced Planck constant, $c$ is the speed of light. The refractive index $n_D = 2.5$ for hBN was chosen at the ZPL energy $E_\text{ZPL}$ of around 4.1 eV.

The changes in the dipole moment upon excitation were evaluated by the time-dependent density functional theory (TDDFT) method using the ORCA code~\cite{neese2018software}. To this end, a cluster model of 120 atoms was described with the cc-pVDZ basis set~\cite{dunning1989gaussian} and the PBE0 density functional~\cite{perdew1996rationale}. The TDDFT results also confirm that the lowest excitations of the 2C and SW defects are essentially between the single pairs of orbitals, which perfectly fits into the scope of $\Delta$SCF method. A justification of $\Delta$SCF method for the 6C defect is provided elsewhere \cite{li2022ultraviolet}.

\section*{Author contribution}
S.L. and A.P. performed the $ab$ $initio$ calculations with a contribution of P.L. and wrote the paper with input from all authors. All authors discussed the results. A.G. conceived the work and supervised his group members.

\section*{Competing interests}
The authors declare that there are no competing interests.

\section*{Data Availability}
The data that support the findings of this study are available from the corresponding author upon reasonable request.

%
%
\begin{acknowledgments}
AG acknowledges the Hungarian NKFIH grant No.~KKP129866 of the National Excellence Program of Quantum-coherent materials project and the support for the Quantum Information National Laboratory from the Ministry of Innovation and Technology of Hungary. A part of the calculations was performed using the KIF\"U high-performance computation units.
\end{acknowledgments}

\bibliography{mainref}

\end{document}


\title{Supplementary Information\\ for \\ Exceptionally strong coupling of defect emission in hexagonal boron nitride to stacking sequences}

\author{Song Li}
\affiliation{Wigner Research Centre for Physics, P.O.\ Box 49, H-1525 Budapest, Hungary}

\author{Anton Pershin}
\affiliation{Wigner Research Centre for Physics, P.O.\ Box 49, H-1525 Budapest, Hungary}
\affiliation{Department of Atomic Physics, Institute of Physics, Budapest University of Technology and Economics,  M\H{u}egyetem rakpart 3., H-1111 Budapest, Hungary}

\author{Pei Li}
\affiliation{Beijing Computational Science Research Center, Beijing 100193, China}
\affiliation{Wigner Research Centre for Physics, P.O.\ Box 49, H-1525 Budapest, Hungary}

\author{Adam Gali}
\affiliation{Wigner Research Centre for Physics, P.O.\ Box 49, H-1525 Budapest, Hungary}
\affiliation{Department of Atomic Physics, Institute of Physics, Budapest University of Technology and Economics,  M\H{u}egyetem rakpart 3., H-1111 Budapest, Hungary}

\maketitle

\section{Supplementary Note 1: Band alignment in different stacking sequences of \MakeLowercase{h}BN} 

To calculate the band alignment diagrams, we used a slab model of 6 layers containing vacuum layer of 15 \AA. We applied a Monkhorst-Pack mesh of $9\times9\times1$ k-points for the Brillouine zone integration. The calculated band gap energies were in close agreement with those of bulk models, see Supplementary Figure~\ref{FigureS1}. We found that $\text{AA}^{\prime}$ and AB stackings exhibit an indirect band gap with the valence band maximum (VBM) at K high symmetry point while the conduction band minimum (CBM) is at M high symmetry point. By contrast, $\text{AB}^{\prime}$ stacking shows a direct band gap, with both VBM and CBM located at K-point.

\begin{figure}
\includegraphics[width=\columnwidth]{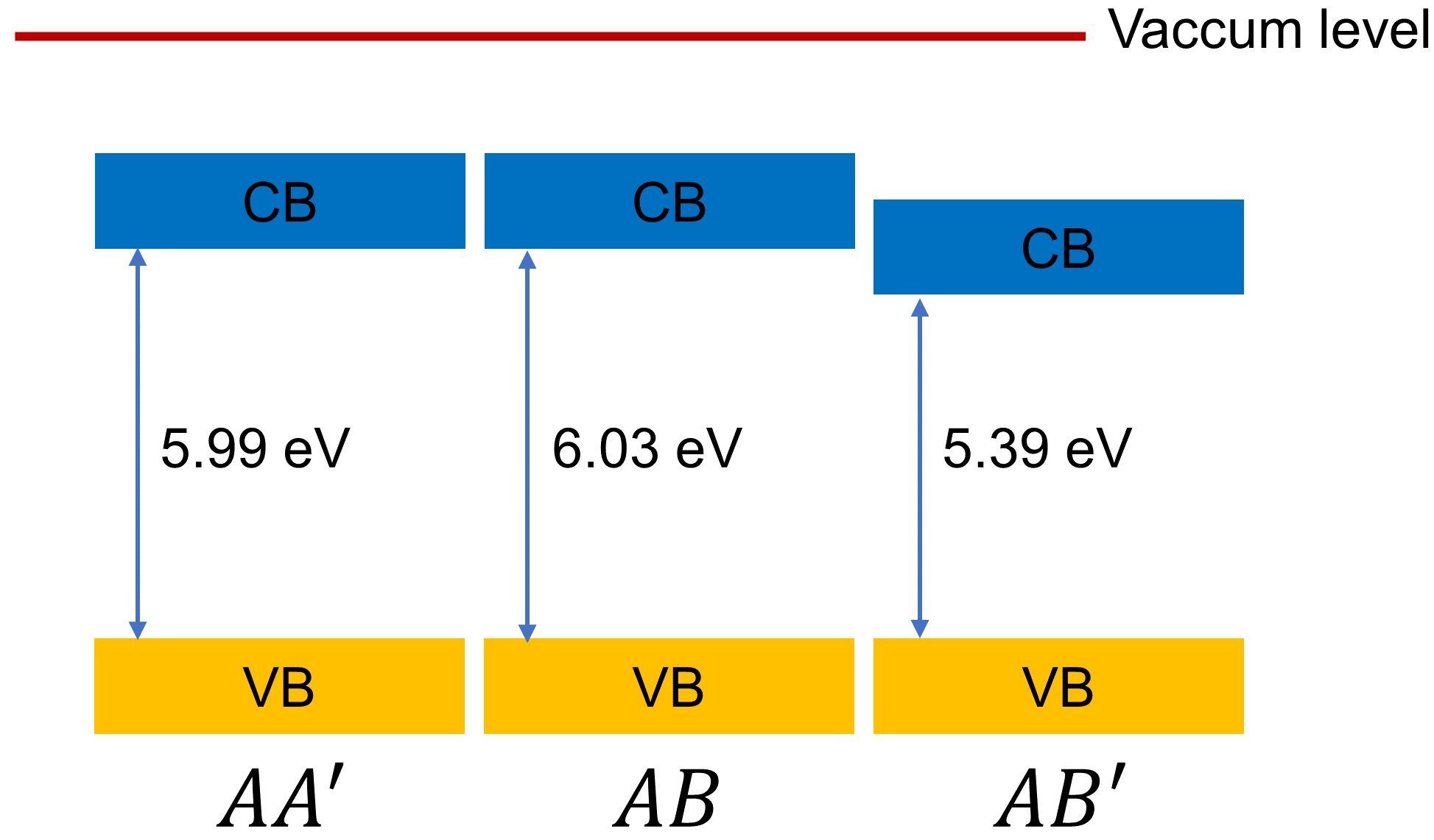}
\caption{\label{FigureS1}%
Band alignment diagram, relative to the vacuum level, of $\text{AA}^{\prime}$, AB and $\text{AB}^{\prime}$ stacking sequences.}
\end{figure}

\begin{figure*}
\includegraphics[width=2\columnwidth]{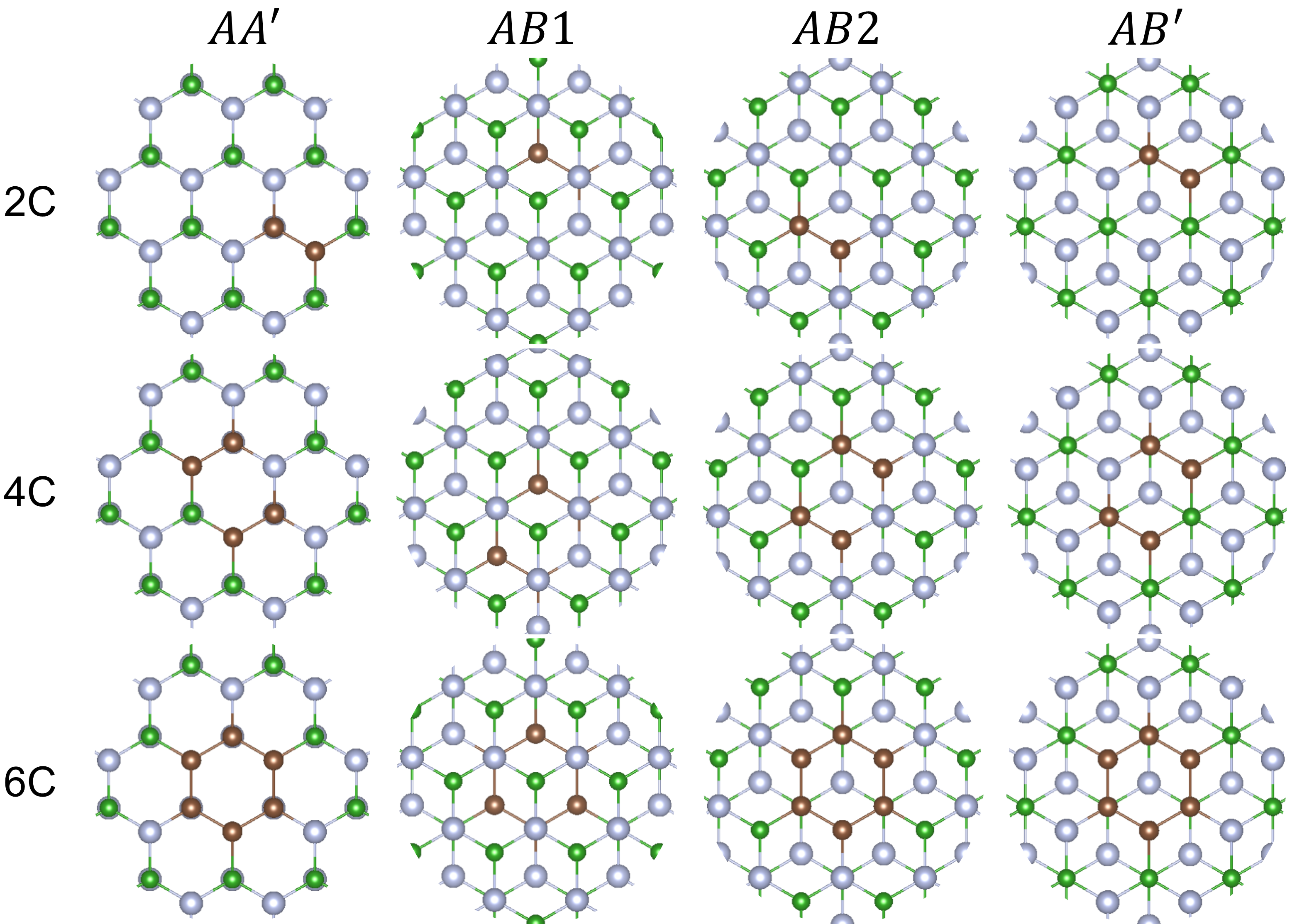}
\caption{\label{FigureS2}%
The carbon defects configurations in four stacking patterns.}
\end{figure*}

\section{Supplementary Note 2: Correction energy for ZPL}

Within the framework of constrained DFT, the singlet excited states cannot be calculated directly: the wavefunction of the singlet excitation is a combination of two Slater determinants, while cDFT method is limited to a single Slater determinant. To overcome this limitation, alternative electronic structure methods, such as CASSCF, DMRG and coupled-cluster methods may be used, yet those are beyond our current scope. Instead, the multi-determinant singlet excited state was constructed based on the Slater's sum rules~\cite{ziegler1991approximate}, given as follows:
\begin{equation}\label{eq:1}
E_{S} = 2E_{S}^{cDFT}-E_{T}^{cDFT} = E_{S}^{cDFT}+E_{corr}\text{.}
\end{equation}
Here, $E_{S}^{cDFT}$ and $E_{T}^{cDFT}$ are the energies of one-electron transitions in the singlet and triplet manifolds, respectively, obtained with the cDFT method. In other words, the multi-determinant $E_{S}$ energy is computed by adding a correction term, $E_{corr}$, to $E_{S}^{cDFT}$, where $E_{corr}$ is defined as follows:

\begin{equation}\label{eq:2}
E_{corr} = E_{S}^{cDFT}-E_{T}^{cDFT}\text{.}
\end{equation}
We note that for carbon defects the correction energy is generally found to be 0.45-0.50 eV, as shown in Supplementary Table~\ref{tab:data}.    

\renewcommand{\arraystretch}{1.5}
\begin{table}[htb]
\caption{\label{tab:data} Calculated correction energies (eV), and ZPL energies without the correction term (eV) using the $9\times9$ supercell of hBN.}
\begin{ruledtabular}
\begin{tabular}{l|cc}
Defect & Correction  & ZPL  \\
\hline
$\text{AA}^{\prime}$-2C & 0.47 & 3.62 \\
$\text{AB1}$-2C & 0.49 & 3.75 \\
$\text{AB2}$-2C & 0.49 & 3.72 \\
$\text{AB}^{\prime}$-2C & 0.43 & 3.57 \\
\hline
$\text{AA}^{\prime}$-4C & 0.47 & 3.84 \\
$\text{AB1}$-4C & 0.52 & 4.04 \\
$\text{AB2}$-4C & 0.50 & 3.96 \\
$\text{AB}^{\prime}$-4C & 0.41 & 3.87 \\
\hline
$\text{AA}^{\prime}$-6C & 0.44 & 3.72 \\
$\text{AB1}$-6C & 0.51 & 3.98 \\
$\text{AB2}$-6C & 0.47 & 3.84 \\
$\text{AB}^{\prime}$-6C & 0.43 & 3.79 \\
\hline
$\text{AA}^{\prime}$-SW & 0.09 & 3.90 \\
$\text{AB1}$-SW & 0.09 & 3.92  \\
$\text{AB2}$-SW & 0.09 & 3.88  \\
$\text{AB}^{\prime}$-SW & 0.07 & 3.90 \\

\end{tabular}
\end{ruledtabular}
\end{table}

\begin{figure*}[hptb]
\includegraphics[width=2\columnwidth]{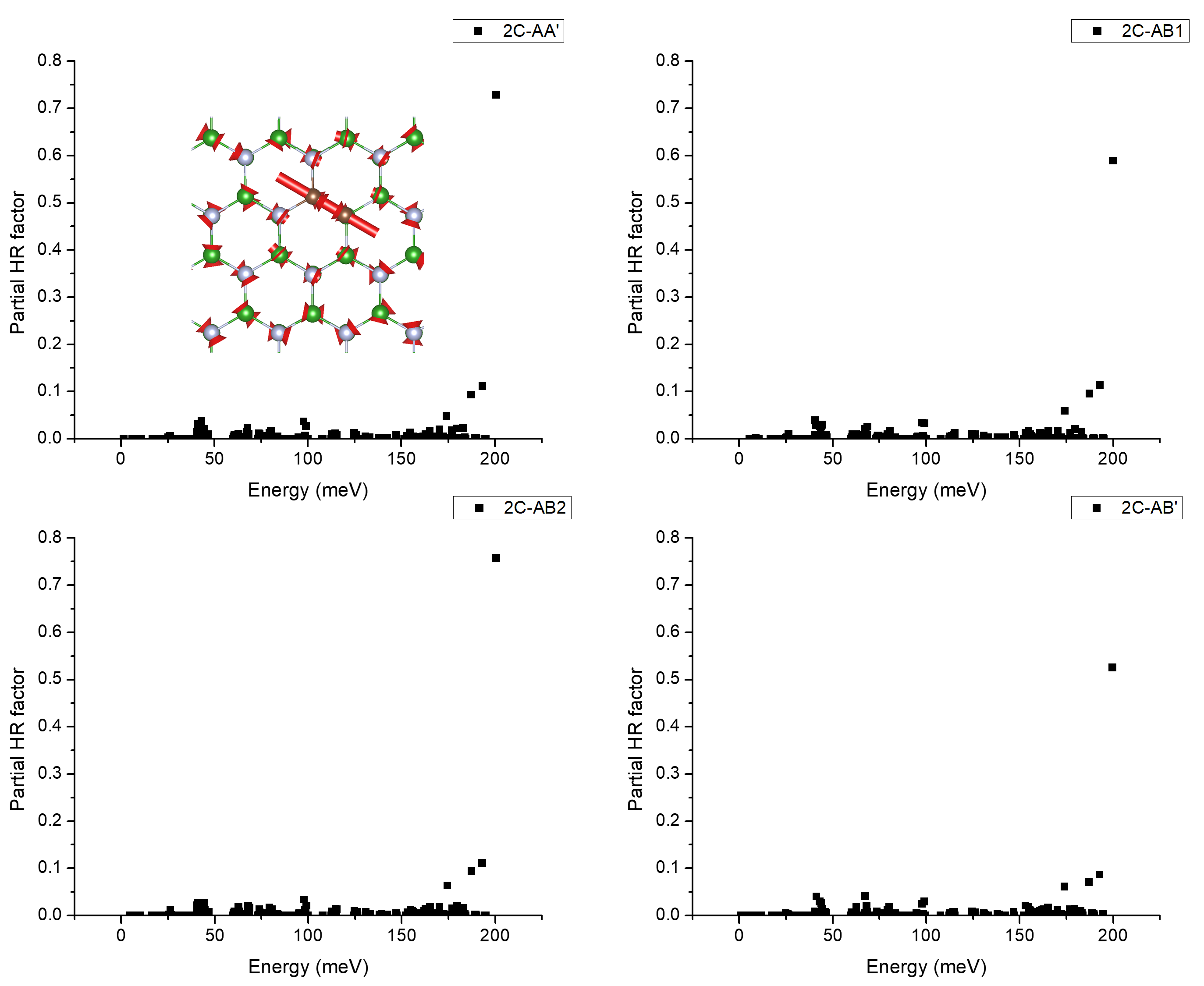}
\caption{\label{FigureS3}%
The partial HR factor distribution of phonon modes during transition from ground state to excited state in 2C defect. The dominant local vibration mode is shown with red arrows.}
\end{figure*}
\begin{figure*}
\includegraphics[width=2\columnwidth]{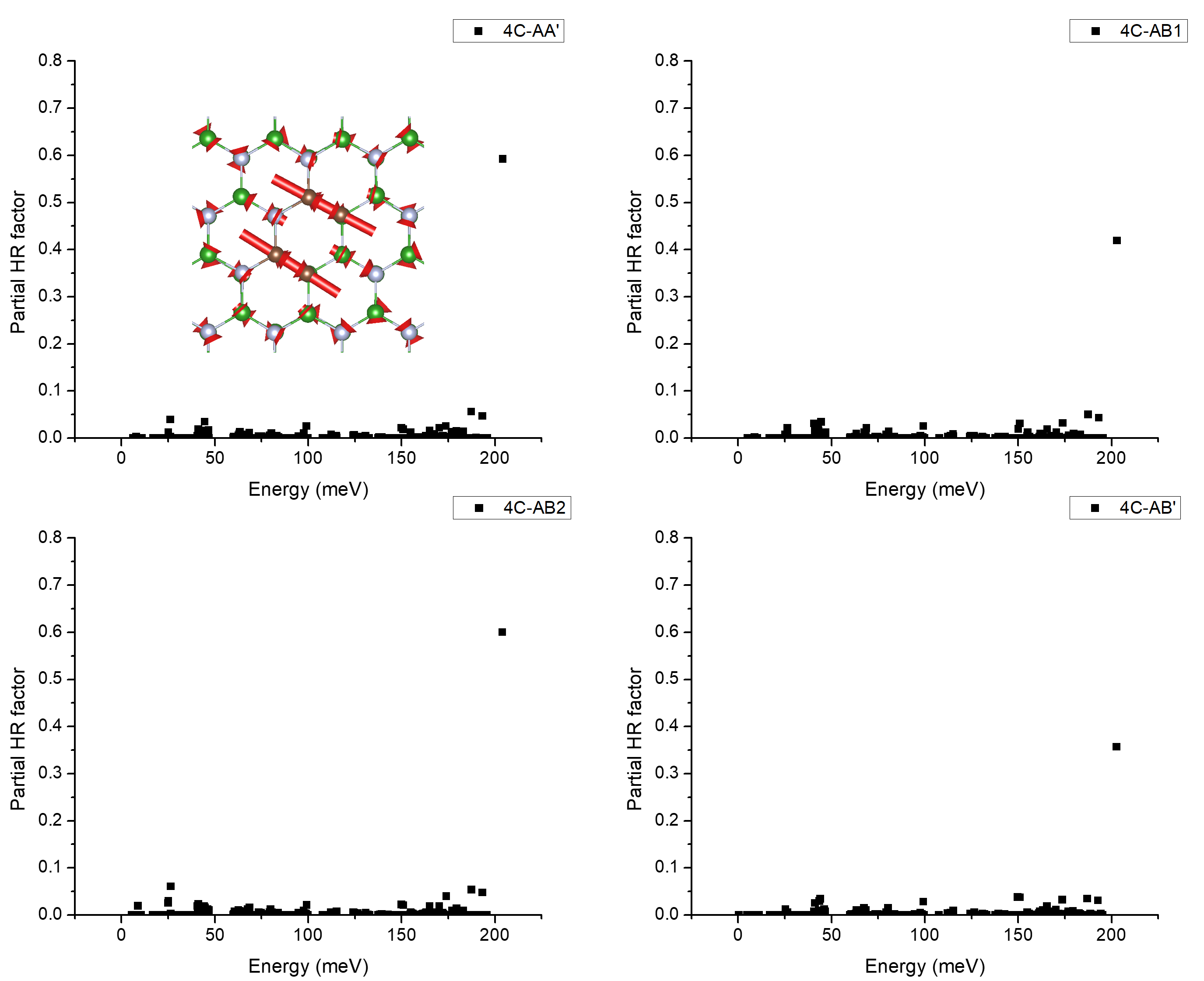}
\caption{\label{FigureS4}%
The partial HR factor distribution of phonon modes during transition from ground state to excited state in 4C defect. The dominant local vibration mode is shown with red arrows.}
\end{figure*}
\begin{figure*}
\includegraphics[width=2\columnwidth]{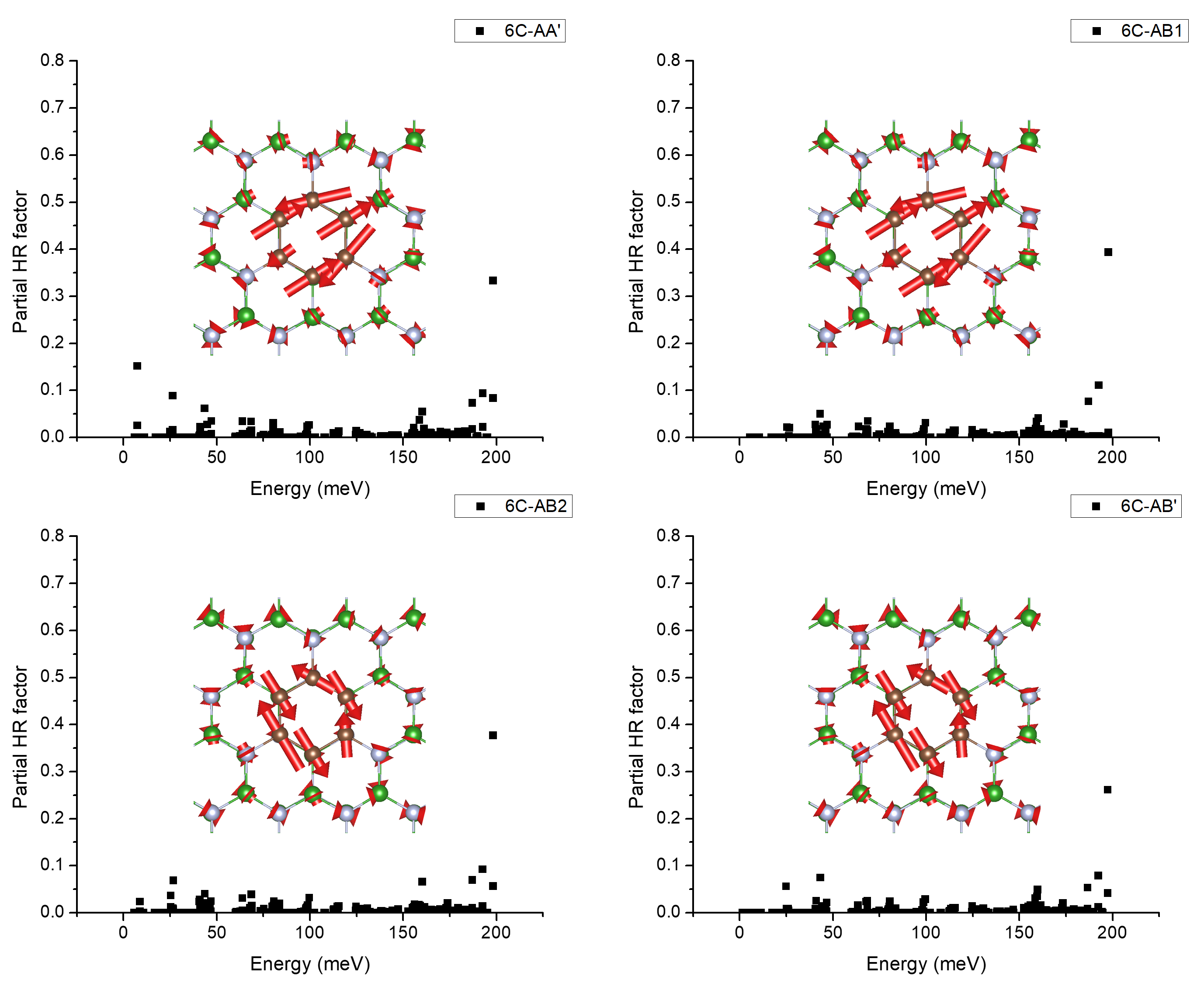}
\caption{\label{FigureS5}%
The partial HR factor distribution of phonon modes during transition from ground state to excited state in 6C defect. The dominant local vibration mode is shown with red arrows.}
\end{figure*}

\begin{figure*}
\includegraphics[width=2\columnwidth]{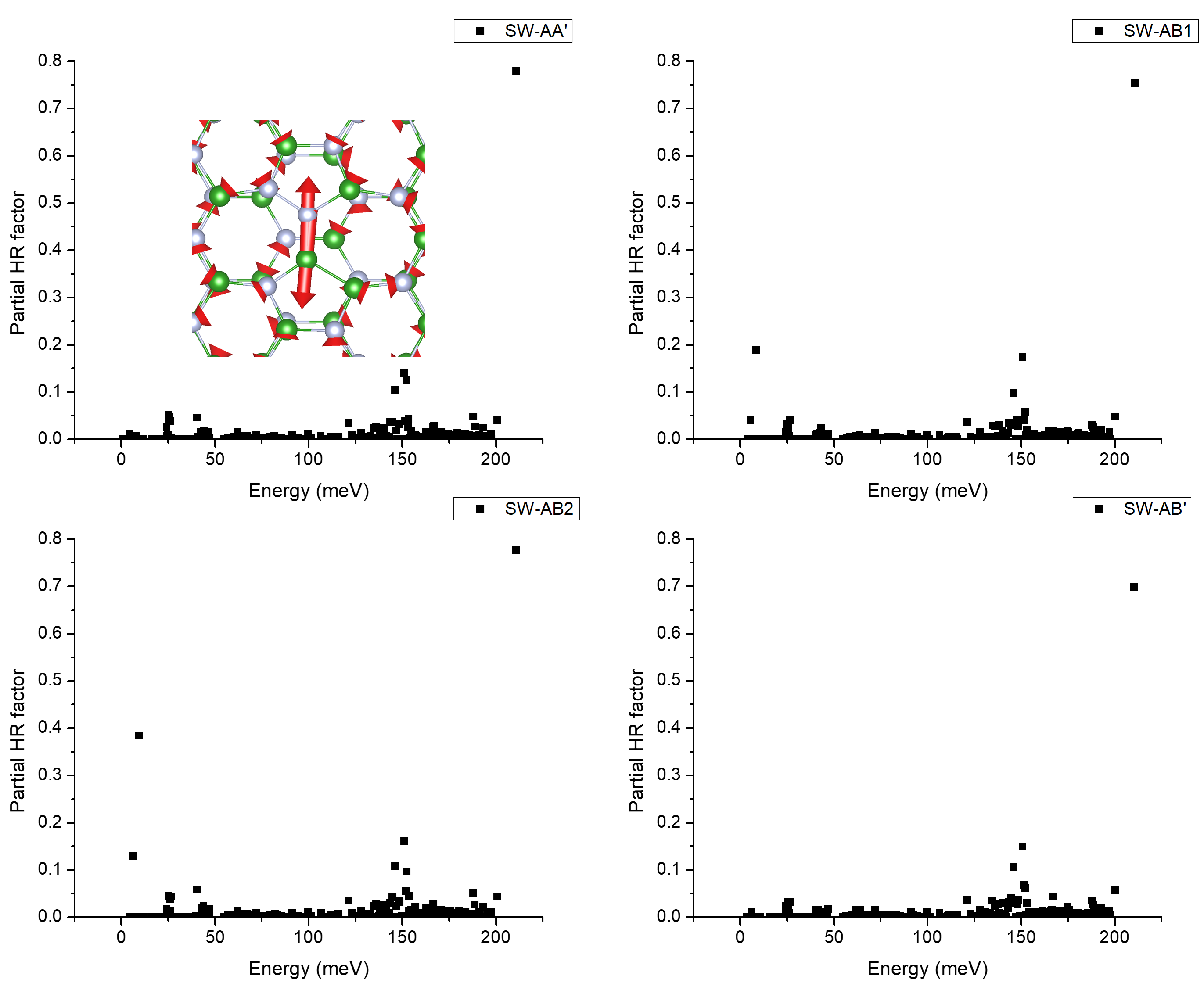}
\caption{\label{FigureS6}%
The partial HR factor distribution of phonon modes during the transition from ground state to excited state in SW defect. The dominant local vibration mode is shown with red arrows.}
\end{figure*}

\begin{figure*}
\includegraphics[width=2\columnwidth]{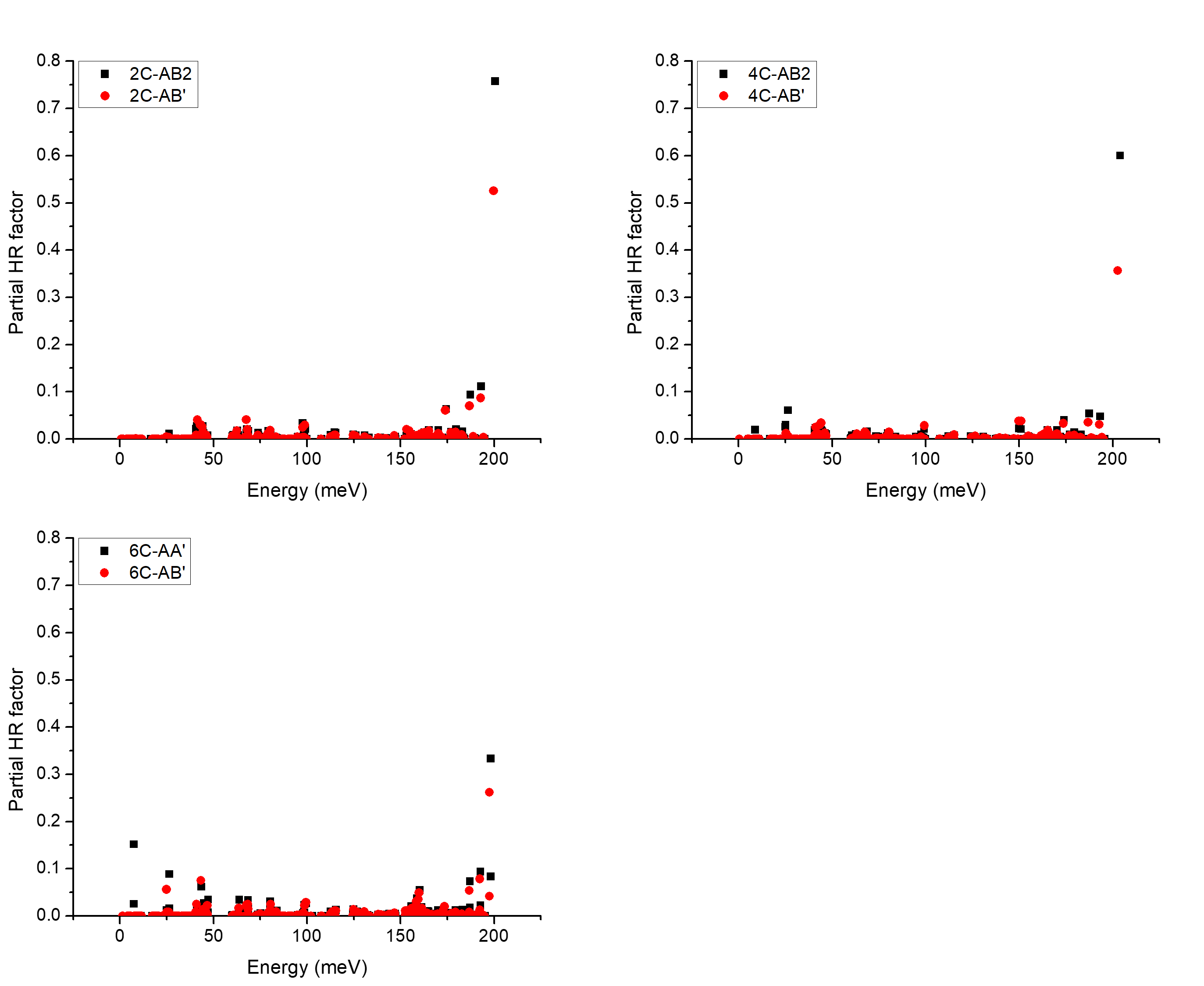}
\caption{\label{FigureS7}%
The partial HR factor distribution of phonon modes during transition from ground state to excited state in different stackings of hBN. For each defect, the stacking sequences leading to the largest change in the HF factors (see Table I) are compared.}
\end{figure*}

\section{Supplementary Note 3: Partial HR factors and local vibration modes}

\begin{figure*}
\includegraphics[width=2\columnwidth]{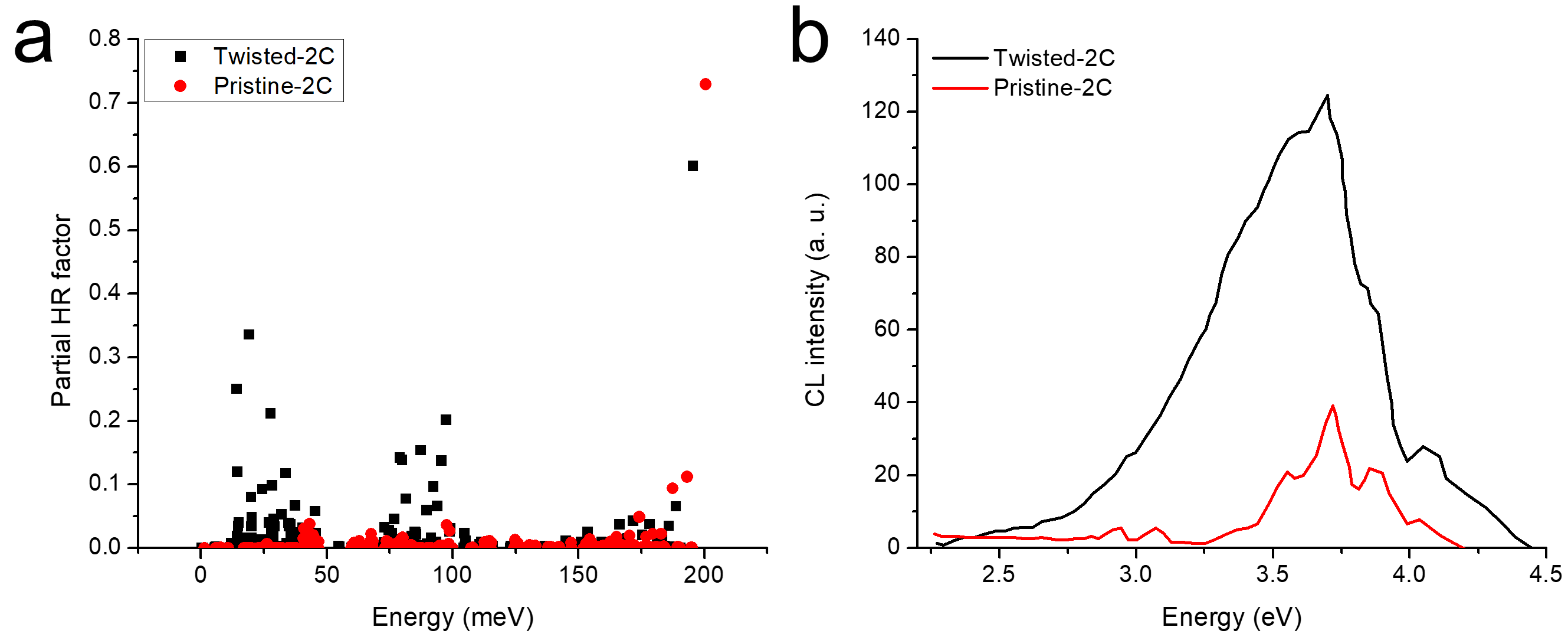}
\caption{\label{FigureS9}%
(a) The partial HR factors of 2C for the optical transition from ground state to the lowest excited state in a twisted bilayer and bulk hBN (AA'). (b) The effect of twisting on the experimental luminescence spectrum of 4.1 eV-emitters. The data is taken from Ref.~\cite{su2022tuning}.}
\end{figure*}

In this section, we plot the partial HR factors for all the defects in four stacking patterns. The partial HR factor is the average number of phonons of a certain mode that are emitted during the optical excitation. In Supplementary Figures~\ref{FigureS3} and ~\ref{FigureS4}, it is easy to identify the stretching mode of the carbon dimer at $\sim$200 meV, which is responsible for the excited state relaxation of 2C and 4C defects regardless of the stacking. Due to the high symmetry of 6C defect, there are two degenerate local vibration modes, as depicted in Supplementary Figure~\ref{FigureS5}. Interestingly, the intensities of the dominant vibration modes vary between different stacking sequences and are suppressed in $\text{AB}^{\prime}$ stacking. This may be related to its narrower band gap, allowing for a stronger interaction between the conduction band and empty defect levels . We also observe a contribution of the low-energy translation modes in $\text{AA}^{\prime}$ stacking which is discussed previously. The contribution of this mode is less apparent in $\text{AB}^{\prime}$ stacking, suggesting that a local sliding at around 6C defect might occur. This translation also happens for SW defect in AB stacking, but is barely visible in $\text{AA}^{\prime}$ and $\text{AB}^{\prime}$ sequences, see Supplementary Figure~\ref{FigureS6}. Direct comparison between the partial HR factors is provided in Supplementary Figure~\ref{FigureS7}. Note that the difference in intensities of the local vibration modes is the major reason for the changes in the photoluminescence spectra. Similarly, Supplementary Figure~\ref{FigureS9} shows the changes in partial HR factors of 2C defect due to twisting along with the respective experimental response.  

\section{Supplementary Note 4: Evolution of the band gap induced by sliding}

We calculate the changes in the band gap energies due to sliding from $\text{AB}^{\prime}$ to $\text{AA}^{\prime}$ stacking, as shown in Supplementary Figure~\ref{FigureS8}. As discussed in the main text, the geometries were not optimized and therefore, the curve is not parabolic. However, the trend shows that the band gap decreases in non-centrosymmetric stackings and reaches its minimum when the other layer is situated on the bridge site, which minimizes the interlayer interaction.

\begin{figure}
\includegraphics[width=\columnwidth]{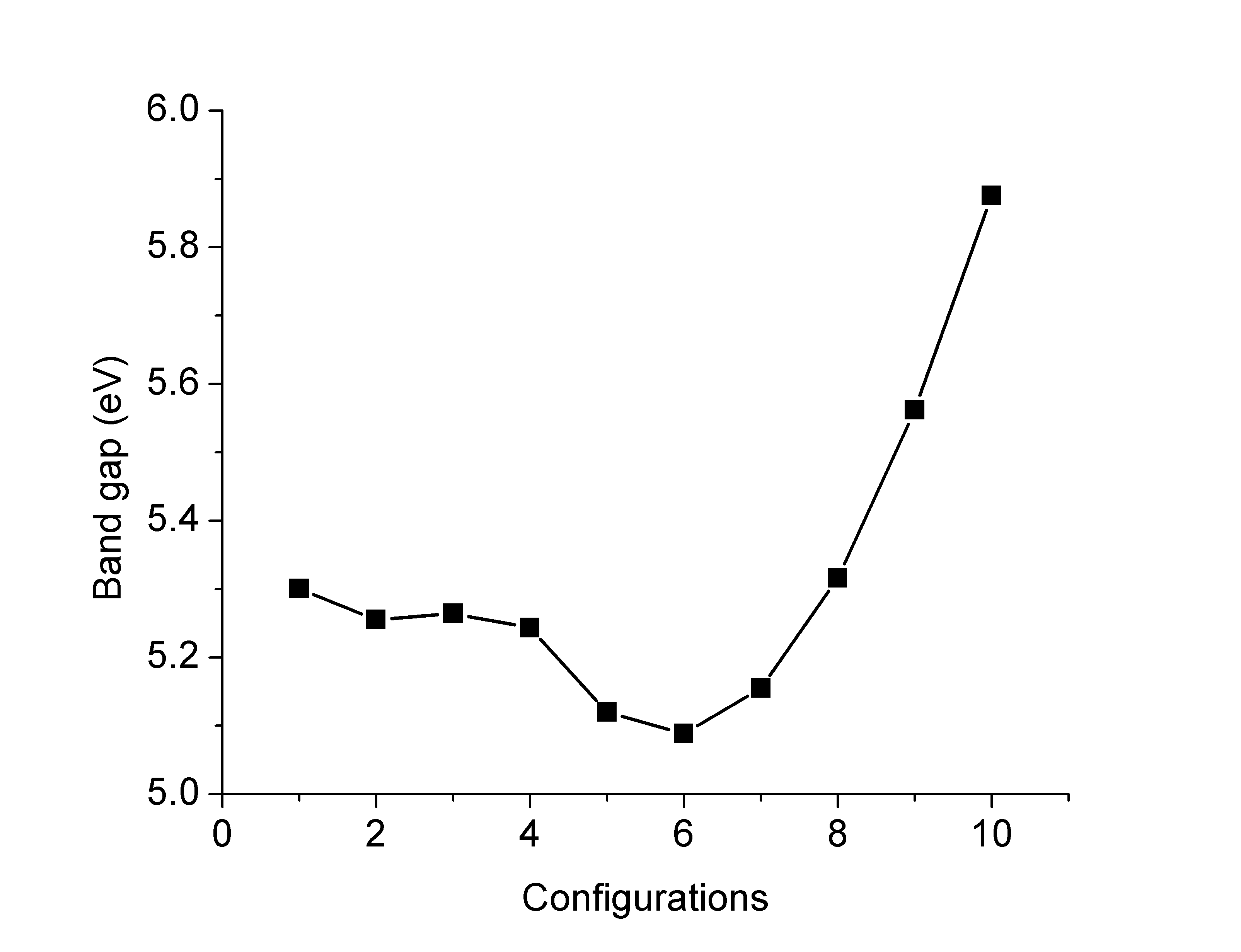}
\caption{\label{FigureS8}%
The band gap evolution during sliding from $\text{AB}^{\prime}$ to $\text{AA}^{\prime}$.}
\end{figure}

\bibliography{mainref}